
\magnification 1200
\baselineskip=14pt
\parskip 10pt plus 1pt minus 1pt
\hsize 16 true cm \vsize 22 true cm
\nopagenumbers
\headline = {\hfill \tenrm\folio\ }
\def\relatif{\ \hbox{{\rm Z}\kern-.4em\hbox{\rm Z}}}
\def\nat{\hbox{{\rm N}\kern-1em\hbox{{\rm I}}}}
\def\reel{\ \hbox{{\rm R}\kern-1em\hbox{{\rm I}}}}
\def\souligne#1{$\underline{\smash{\hbox{#1}}}$}
\def\1op{\ \hbox{{\rm 1}\kern -0.23em\hbox{{\rm I}}}}
\def\laeq{\ \raise .3ex\hbox{{\rm {$<$}}\kern-.8em\lower 1.3ex\hbox{{\rm
{$\sim$}}}}\ }

\ \
\vskip 2 cm
\centerline{\bf
  Quasi Exactly Solvable 2$\times$2 Matrix  Equations}
\vskip 1.5 true cm

\centerline {\bf Y. Brihaye}
\centerline {Facult\'e des Sciences, Universit\'e de Mons-Hainaut}
\centerline {B-7000 Mons, Belgium}
\vskip .5 cm
\centerline {\bf P. Kosinski$^{\dagger}$}
\centerline {Department of Theoretical Physics, Lodz University}
\centerline {90-236 Lodz, Poland}

\vskip 3 cm
\centerline{\bf
       Abstract}
\par \noindent
We investigate the conditions under which systems of two
differential eigenvalue  equations are quasi exactly solvable.
These systems reveal a rich set of algebraic structures.
Some of them are explicitely described. An exemple of
quasi exactly system is studied which provides a direct
counterpart of the Lam\'e equation.

\vskip 3 cm
\noindent {Mons-Preprint} \hfill\break
\noindent {July 1993} \hfill \break
\noindent {$^{\dagger}$ Work supported by the Grant N$^0$
 ERB-CIPA-CT-93-0670 of CEC}
\vfil \eject

 \noindent 1. \souligne {\bf Introduction}

Quasi exactly solvable (QES)  equations are differential, spectral equations
(typically Schr\"odinger eigenvalue equations) for which  a finite number of
eigenvalues and eigenvectors can be found by algebraic methods.
In this respect, they contrast with the so called
completely solvable equations
(like the quantum harmonic oscillator or the Coulomb problem)
where the whole discrete  spectrum can be computed algebraically.

After their introduction, a few years ago, by A.~V.~Turbiner [1]
and A.~G.~Ushveridze [2], the quasi exactly solvable equations received
many new developments (see Refs.~[3,4] for reviews).

Scalar  QES equations in one  variable are now well understood [1-5].
The basic ingredient is that, in a good variable, the linear
differential operator defining them can be expressed in terms of
a projectivized representation of the generators of the group
SL(2,$\reel$ \ \ ).
Scalar equations in two variables, one real and one Grassmann,
were investigated in Refs. [3,4,6]. All results obtained in this
context can be reformulated in terms of systems of two equations
in one real variable. The relevant algebraic structure behind this
type of  equations is the graded algebra osp(2,2).

In this paper, we want to revisit the systems of two coupled
QES equations from a more general point of view. The understanding
of these equations requires the knowledge of all 2$\times$2
matrix differential operators that leave invariant the doublets
of polynomials of given degrees, say $m$ and $n$. Such an investigation
is reported in Sect. 2, where we put the main emphasis on the
algebraic aspects of these sets of differential operators.

In Sect.3 we describe the possible forms of  systems of two
quasi exactly and completely solvable equations.
We compare the number of free parameters they can depend on.
Finally, in Sect.4 we present an explicit example of such system.
It is related to the stability of sphaleron
solutions in the two dimensionnal Abelian-Higgs model.

Before to start with the
systems, we briefly  recall the basic features of
the scalar QES equations in one real variable.
They are strongly  related to
the  linear, differential operators that preserve ${\cal P }_n$:
the space of polynomials of degree $n$ in a variable, say $x$.
Among these operators, the three given by
$$  J_n^+ = x^2{d\over {dx}} - n x \ \ ,
\ \ J_n^{0} = x{d\over {dx}} -{n\over 2}\ \ \ ,
\ \ J_n^- = {d\over {dx}} \eqno(1)$$
play a considerable role. They obey the commutation relations of
the  generators of the group  SL(2,$\reel$ \ \ ):
$$ [ J_n^+ , J_n^-] = -2 J_n^0 \ \ \ , \ \ \
   [ J_n^{\pm} , J_n^0] = \mp  J_n^{\pm}  \ \ \ , \ \ \
    \eqno (2)$$
and therefore constitute an irreducible  representation
of dimension $n+1$ of the algebra of this group.
Using the irreductibility of the representation
it is easy to show that the linear
differential operators preserving the space ${\cal P}_n$
are the elements of the envelopping algebra generated
by the three operators (1). Accordingly, if $P$ denotes a
polynomial in three variables, then the eigenvalue equation
$$  P(J^+_n, J^0_n, J^-_n) \cdot F(x) = \lambda F(x)     \eqno(3)  $$
is equivalent to a system of $n+1$ algebraic equations.
If $P$ is quadratic, it is possible, formally, to
rephrase Eq. (3) as a Schr\"odinger equation by mean of
a change of variable and of a change of function.

 \noindent 2. \souligne {\bf Algebraic Structure of QES Systems}

The purpose of this section is to describe  the algebraic features of
quasi exactly solvable  systems of two equations.
We denote by ${\cal P}_{m,n}$ the space
of doublets of polynomials in $x$, the first (resp. second)
component being of degree $m$ (resp.$n$).
Given the integers $m,\ n$, we will   consider the
set of operators defined by
$$ \eqalign{
&T^+ = \pmatrix {J^+_{m} &0 \cr 0 &J^+_n\cr}\ ,\
 T^0 = \pmatrix {J^0_{m} &0 \cr 0 &J^0_n\cr}\ ,\
 T^- = \pmatrix {J^-_{m} &0 \cr 0 &J^-_n\cr} \ ,\cr
&  J=  {1\over 2} \pmatrix {n+ \Delta &0 \cr 0 &n\cr}\ \ \
, \ \ \Delta \equiv n-m  \ \ \ \ \     \cr
&\bar Q_{\alpha} =
 \Bigl( \prod_{j=0}^{\Delta - \alpha} (x{d\over dx}-(n+1-\Delta)-j) \Bigr)
   ({d\over {dx}})^{\alpha -1}  \sigma^+  \ \ \ ,
          \ \ \alpha = 1,2,\cdots \Delta + 1    \cr
&Q_{\alpha} = x^{\alpha-1} \sigma^-  \ \ \ \ ,
             \ \ \alpha = 1,2,\cdots \Delta + 1   \cr } \eqno(4)$$
It is easy to show that these operators are linearly independent
and that each of them
preserves the space  ${\cal P}_{m,n}$. Moreover,
reasoning component by component, one can convince
that any linear operator preserving  ${\cal P}_{m,n}$
can be constucted polynomialy from the 6+2$\Delta$ generators  (4).
So, the operators (4) play for systems of two equations
the same role as the operators (1) do for scalar equations

We want to demonstrate that the operators (4) close under
(anti-) commutation, independently of $m$ and $n$.
The subalgebra sustended by $T^0, T^{\pm}$ is isomorphic to the
algebra of  SL(2,$\reel$ \ \ ). The SU(2) generators, say $T_k$ can be
recovered by the combinations
$$  T_1 = {1\over 2} (T^- -T^+) \ \ ,\ \
    T_2 = {i\over 2} (T^- +T^+) \ \ , \ \
    T_3 = T^0  \eqno(5)$$
The commutation relations and the  Casimir operator associated to
this (reducible) representation are respectively
$$  [T_j, T_k] = i \epsilon_{j k l} T_l   \eqno(6)$$
$$  C \equiv T_1^2 + T_2^2 + T_3^2  =
 {1 \over 4} \pmatrix { m(m+2) &0 \cr 0 &n(n+2)\cr} \eqno (7) $$


The operators (4) behave differently under the
 commutation with the matrix $J$
(whose form in (4) is defined for later convenience):
 $$ [J , T_k] = 0 \ \ \ , \ \ \
   [J , Q_{\alpha}] = - {\Delta \over 2} Q_{\alpha} \ \ \ , \ \ \
   [J , \bar Q_{\alpha}] =  {\Delta \over 2} \bar Q_{\alpha}  \eqno(8)$$
{}From now on we will refer to
$T_a, J$ as to bosonic operators and to  $Q$
 (resp. $\bar Q$) as to fermionic (resp. anti-fermionic) operators.
The $J$-weight of bosonic operators is zero while that of fermionic
(resp. antifermionic) ones is $-\Delta /2$ (resp.$+\Delta/2$).

For the commutation rules between bosonic and fermionic operators we obtain
$$ \eqalign {
   &[Q_{\alpha} , T^+] = (1- \alpha + \Delta) Q_{\alpha + 1} \cr
   &[Q_{\alpha} , T^0] = (1- \alpha + {\Delta \over 2}) Q_{\alpha} \cr
   &[Q_{\alpha} , T^-] = (1- \alpha ) Q_{\alpha - 1} \cr } \eqno(9a)$$
$$ \eqalign {
  &[\bar Q_{\alpha} , T^+] = -(1- \alpha ) \bar Q_{\alpha-1} \cr
   &[\bar Q_{\alpha} , T^0] = -(1- \alpha + {\Delta \over 2}) \bar Q_{\alpha} \cr
   &[\bar Q_{\alpha} , T^-] = -(1- \alpha + \Delta) Q_{\alpha + 1} \cr
   } \eqno(9b)$$
demonstrating that the sets of
 $Q_{\alpha}$'s and $\bar Q_{\alpha}$'s
 transform according to the
$s = {\Delta \over 2}$ representation  of the  SL(2,$\reel$ \ \ )
subalgebra.
This statement becomes more transparent if we redefine these operators
according to
$$  \hat Q_{\rho} =
 \pmatrix {2s \cr s +\rho \cr}^{1/2}
 Q_{s+1+\rho} \ \ \ , \ \ \
       -s \leq \rho \leq s \ \ \ , \ \  s\equiv {\Delta \over 2} \eqno(10)$$
\par The computation of the anti-commutation relations
between fermionic and anti-fermionic operators
is more involved. Formally,  we obtain
$$\eqalign {\lbrace \bar Q_{\alpha}, Q_{\beta} \rbrace &=
{\cal M}_{\alpha \beta} (T^-)^{\alpha-\beta}\ \ \ ,
{\rm if} \ \alpha \geq \beta\cr
&= (T^+)^{\beta-\alpha} {\cal M}_{\alpha\beta}\ \ \ ,
{\rm if} \ \alpha\leq \beta\cr} \eqno(11)$$
with
$${\cal M}_{\alpha\beta} =
 \prod^{\Delta-\alpha}_{j=0} (T^0+J_c - j \1op - (\beta-1) P_2) \
 \prod^{\beta-2}_{k=0} (T^0+J- k \1op - (\Delta + 1-\alpha) P_1) \eqno(12)$$
and
$$J_c  = (\Delta - 1) \1op - J\ ,\ P_1 = \pmatrix {1 &0\cr 0&0\cr}\ ,
\ P_2 = \pmatrix {0 &0\cr 0 &1\cr} \eqno(13)$$
In the definitions (12) it is understood that the first (resp. second)
product is $\1op$ if $\alpha \geq \Delta$ (resp. if  $\beta \leq 2$).
The form  (11) is
not satisfactory because the projectors $P_1,
P_2$ entering in  ${\cal M}_{\alpha\beta}$ are not
expressible in terms of $J$ and of $C$ (the Casimir)
in an $n$-independent way. These matrices  however obey  the identity
$$(J+({1-\Delta\over 2}))P_2 = -{1\over {2\Delta}}
 (J^2+(1-2\Delta) J + \Delta(\Delta-1)-C)    \eqno(14)$$
\par In order to demonstrate that
 ${\cal M}_{\alpha\beta}$ is independent on $n, P_1, P_2$
 we write formally
$${\cal M}_{\alpha\beta} = A_{\alpha\beta} P_2 + B_{\alpha\beta} \eqno(15)$$
Using the fact that  $P_1,P_2$ are projectors,
a direct evaluation of the matrix  $A_{\alpha\beta}$ yields
$$A_{\alpha\beta}
    = \prod^{b_1}_{j=a} (T^0+X-j) \prod^{c}_{j=b_1 + 1}(T^0 - X-j) -
      \prod^{b_2}_{j=a} (T_0-X-j) \prod^{c}_{j=b_2 + 1}(T_0+X-j)
\eqno(16) $$
with
$$X\equiv J + ({1-\Delta\over 2})   \eqno(17)$$
and
$$ a \equiv {1-\Delta\over 2} , c = \beta - \alpha + {\Delta-1\over 2}
\ ,\ b_1 = -\alpha + {\Delta+1\over 2}\ , \ b_2 = \beta-{\Delta+3\over 2}
\eqno(18)$$
{}From Eq. (16), it is clear that the matrix $A$ contains $X$
as  a factor. Then  the use  of (16)
together with the identity (14) demonstrates the statement
that ${\cal M}$ are independent on $n$ and on the $P$'s.
For completeness we also  mention  the  relations
  $$  \lbrace Q_{\alpha} , Q_{\beta} \rbrace = 0 \ \ \ \ ,
       \lbrace \bar Q_{\alpha} , \bar Q_{\beta} \rbrace = 0 \eqno(19) $$

So,  Eqs.(9)-(19) finally  show  that the 6+2$\Delta$ generators (4)
close under (anti)-commutation to a family of operator
algebras indexed by the integer $\Delta$.
 In order to find the abstract algebras for which the above
operator algebras provide the representations
we  have to check whether  all Jacobi identities are
fullfilled on abstract level and,
if not,  to modify some of the commutation rules.

It appears that  only the Jacobi identities for
$Q,Q$ and $\bar Q$ or for $\bar Q \bar Q$ and $Q$ impose further constraints
on generators. This suggests that just the  anticommutation rules (19)
should be changed.
The anticommutator $\lbrace Q_{\alpha},Q_{\beta} \rbrace$ has weight
2 with respect to $J$. Therefore its value should also  be
quadratic in $Q$'s (we exclude the more complicated objects like
$Q Q Q \bar Q$ etc.). By analysing the Jacobi identity for
$Q_{\alpha}, Q_{\beta}, \bar Q_{\gamma}$ we arrive at the conclusion
that the coefficients of the quadratic form on the right hand side
cannot depend on $T_i$'s and on $J$; i.e. they have to be
numerical. Therefore the only possibility is that we decompose
 $\lbrace Q_{\alpha},Q_{\beta} \rbrace$ into SU(2) irreducible
components and keep only some of them on the right hand side.
We solved this problem for $\Delta=$ 2 and 3.
It appeared  that in order to satisfy the Jacobi identity
it is sufficient to keep only the lowest non vanishing representation.
The resulting relation can be viewed as a quadratic constraint
on the $Q$'s. After this constraint is imposed, no further constraint
is implied by the Jacobi identities.

We now collect the cases where an abstract algebra was found.

\noindent  Case $\Delta = 0$

\noindent
In the case $\Delta = 0$, we have an ambiguity in defining the algebra.
We can treat it according to the general scheme above (in this case
$J\propto \1op$). Alternatively, we can also put $J = \sigma_3$ and impose the
{\it commutation} rule on $Q, \bar Q$. In this case one obtains the
the Lie algebra of  SL(2,$\reel$ \ \ )$\times$SL(2,$\reel$ \ \ )

\noindent  Case $\Delta = 1$

\noindent
This case was considered at lenght in Refs. [3,4,6]. The generators
(4) constitute a representation of the graded algebra osp(2,2).
The  anticommutators of $Q$ and $\bar Q$
close into linear combinations of  $T^a$ and $J$
(see  Eq.(9)).
So, the relevant algebra is still a Lie (super) algebra; this
is not the case anylonger when $\Delta > 1$

\noindent  Case $\Delta = 2$

\noindent
The algebra associated with the case $\Delta = 2$  is better
handled when we express the operators $Q_{\alpha}$ in the vector basis,
i.e.

$$ \eqalign {
& \Theta_1 = { Q_3 - Q_1 \over 2}  \ \ , \ \
  \Theta_2 = { Q_3 + Q_1 \over 2 i} \ \ , \ \
  \Theta_3 = - Q_2   \cr
& \bar \Theta_1 = {\bar Q_3 - \bar Q_1 \over 2}  \ \ , \ \
  \bar \Theta_2 = - {\bar Q_3 + \bar Q_1 \over 2 i} \ \ , \ \
  \bar \Theta_3 = \bar  Q_2   \cr } \eqno (20) $$
In this basis, the relations (8),(9),(10) become respectively
$$ [ J , \Theta_j] = - \Theta_j \ \ \ , \ \ \
     [ J , \bar\Theta_j] =  \bar \Theta_j  \eqno(21)$$
$$ [ T_i , \Theta_j ] = i \epsilon_{i j k} \Theta_k  \ \ \ , \ \ \
   [ T_i , \bar \Theta_j ] = i \epsilon_{i j k} \bar\Theta_k \eqno(22) $$
$$ \lbrace \Theta_i , \bar \Theta_j \rbrace =
      \delta_{i j} ( {C \over 6} + {J(J+1) \over 2})
     -{1\over 2} ( T_i T_j + T_j T_i - {2\over 3} \delta_{i j} C)
     + i \epsilon_{i j k} (J - {1\over 2}) T_k  \eqno(23)$$
In the last relation, the right hand side is decomposed according
to irreducible representations of SU(2).
Using such a decomposition for the anticommutators (19)
and keeping only  the scalar part on the right hand side, we obtain instead
$$ \lbrace \Theta_j , \Theta_k \rbrace = {2\over 3}
       \delta_{j k}
 (\Theta_1^2 + \Theta_2^2 + \Theta_3^2)  \eqno(24) $$
and a similar relation for $\bar \Theta$. We checked that all the Jacobi
identities are fullfilled with the choice (24).

\noindent  Case $\Delta = 3$

\noindent In this case, we found it convenient to express the operators
$Q_{\alpha}, \bar Q_{\alpha}$
 ($\alpha = 1,2,3,4$) in a  spin-tensor basis. It consists
in  linear combinations of the $Q$, say
$Q_{i a}$, where the index $i=1,2,3$ (resp.  a=1,2) transform
according to the $s=1$ (resp. $s=1/2$) representation of SU(2).
The components of the spin-tensor $Q_{i a}$ read

$$\eqalign {
&Q_{3{1\over 2}} = \sqrt {2} Q_3\ \ , \ \ \bar Q_{3{1\over 2}} = \sqrt {2}
\bar Q_3\cr
&Q_{3{1\over 2}} = \sqrt {2} Q_2\ \ , \ \ \bar Q_{3-{1\over 2}} = -\sqrt {2}
\bar Q_2\cr
&Q_{1{1\over 2}} = {1\over {\sqrt {2}}} (Q_2-Q_1) \ \ \ , \ \ \
\bar Q_{1-{1\over 2}} = {-1\over {\sqrt {2}}}
(\bar Q_2-\bar Q_4)\cr
&Q_{1-{1\over 2}} = {1\over {\sqrt {2}}} (Q_1-Q_3) \ \ \ , \ \ \
\bar Q_{1-{1\over 2}} = {1\over {\sqrt {2}}}
(\bar Q_1-\bar Q_3)\cr
&Q_{2-{1\over 2}} = {i\over {\sqrt {2}}} (Q_2+Q_4) \ \ \ , \ \ \
\bar Q_{2{1\over 2}} = { i\over {\sqrt {2}}}
(\bar Q_2-\bar Q_4)\cr
&Q_{2-{1\over 2}} = {i\over {\sqrt {2}}} (Q_1+Q_3) \ \ \ , \ \ \
\bar Q_{2-{1\over 2}} = {-i\over {\sqrt {2}}}
(\bar Q_1-\bar Q_3)\cr } \eqno (25)$$

and  obey the identities
$$ Q_{i a}(\sigma_i)_{a b} = 0\ \ \ ,\ \
 (\sigma_i)_{a b} \bar Q_{i b} = 0  \eqno(26)$$
which guarantee that only the  components corresponding to
the   representation $s=3/2$ are selected.

The advantage of the spin tensor basis is that
 the commutators (9) write in a manifestly  covariant way:
$$ \eqalign {
&[T_i,Q_{j a}] = i\epsilon_{ijk}Q_{k a} + {1\over 2}
Q_{j b} (\sigma_i)_{ b a}\cr
&[T_i, \bar Q_{j a}] = i\epsilon_{ijk} \bar Q_{k a}
- {1\over 2} (\sigma_i)_{a b} \bar Q_{j b}\cr} \eqno(27)$$
as well as the anticommutators (11):
$$\eqalign {\lbrace \bar Q_{i a}, Q_{j b}\rbrace
&= 2 T_i(T \cdot \sigma)_{a b} T_j - {2\over 3} C\delta_{ij}
(T \cdot \sigma)_{a b}\cr
&+ {i\over 3} \epsilon_{ijk} T_k C \delta_{a b}-{1\over 3}
C(T_i\sigma_j+T_j\sigma_i)_{a b}\cr
&+ 2(J-1) (T_iT_j-{2\over 3} \delta_{ij}C)\delta_{a b}\cr
&+ i(J-{3\over 2})\epsilon_{ijk} \lbrace T_k, T_{\ell}\rbrace
(\sigma_{\ell})_{a b}\cr
&+{i\over 2} (\lbrace T_j, T_k\rbrace \epsilon_{ik\ell}-\lbrace T_i,
T_k\rbrace \epsilon_{jk{\ell}}) (\sigma_{\ell})_{a b}\cr
&-{i\over 3} (J-1)\epsilon_{ijk} C(\sigma_k)_{a b}\cr
&+{i\over 3} (J-2) (5J-3)\epsilon_{ijk} T_k \delta_{a b}\cr
&+{1\over 3} J(J-2) (T_i(\sigma_j)_{a b}+T_j(\sigma_i)_{a b}
                     -4 \delta_{ij}(T \cdot \sigma)_{a b}) \cr
&+{1\over 3} J(J-1)(J-2) (i \epsilon_{ijk} (\sigma_k)_{a b}
                          -2 \delta_{ij}\delta_{a b} ) \cr} \eqno(28) $$

Keeping again the lowest dimensional representation
appearing in the decomposition of anticommutator
$ \lbrace  Q_{i a}, Q_{j b}\rbrace$ (in this case it is the
vector representation because the scalar piece identically vanishes), i.e.
$$ \eqalign {
\lbrace Q_{i a} , Q_{j b} \rbrace =
 &{1\over 2} (\sigma_2)_{c d} \lbrace Q_{i d} , Q_{j c} \rbrace
              (\sigma_2)_{a b}
  + {2 \over 5} \delta_{i j}  \lbrace Q_{k a} , Q_{k b} \rbrace \cr
 &- {i \over 20}
 \Bigl( ( \sigma_2 \sigma_i)_{a b} \epsilon_{j k l} +
      ( \sigma_2 \sigma_j)_{a b} \epsilon_{i k l} \Bigr) (\sigma_2)_{c d}
         \lbrace Q_{k d} , Q_{k c} \rbrace \cr } \eqno(29)$$
we have checked that all Jacobi identities are fullfilled.


\noindent 3. \souligne {\bf {Systems of QES equations}}

\par The 2$\times$2 differential matrix operators which are quasi exactly
solvable can be expressed in terms of elements of the envelopping algebras
constructed from the generators (4).  Using  the same notations as
in Ref.[6], these operators have the following form
$$T_k (x) = \sum^k_{i=0} a_{k,i}(x) {d^i\over {dx^i}}   \eqno(30)$$
where the coefficient $a_{k,i}(x)$ are 2x2 matrices
with polynomial entries. Their degrees are indicated here in
the  square brackets
$$a_{k,i}(x) =
\pmatrix {A^{[k+i]}_{k,i} &B^{[k+i-\Delta]}_{k,i}\cr
C^{[k+i+\Delta]}_{k,i} &D^{[k+i]}_{k,i}\cr}          \eqno(31)$$
\par Off course not all the polynomials $A_{k,i}$,$B_{k,i}\cdots $
 are arbitrary: for $k\geq \Delta$ and for generic $n$, the
most general  operator  $T_k(x)$ depends on
$$ 4(k+1)^2 \ \ \  , \ \ \  ((k+1)^2)   \eqno(32) $$
arbitrary parameters, independly on  $\Delta$
(the corresponding number
 for scalar equations is given in parenthesis).
For $k< \Delta$ the situation is more complicated because some
constraints may become dependant.

\par For the operator $T_k(x)$ to be exactly solvable, the degree of the
polynomials in $a_{k,i}$ need to be  as follows
$$a_{k,i}(x) =
 \pmatrix {A^{[i]}_{k,i} &B^{[i-\Delta]}_{k,i}\cr
           C^{[i+\Delta]}_{k,i} &D^{[i]}_{k,i}\cr}   \eqno(31')$$
they are not constrained, so that these equations depend on
$$ \eqalign {
&2(k+1)(k+2) + {\Delta(\Delta-1)\over 2}\ \ \ {\rm for} \ \ \ k \geq \Delta
\cr
&{3 \over 2}(k+1)(k+2) + (k+1)\Delta \ \ \  {\rm for} \ \ \ k < \Delta \cr}
  \eqno(32')$$
free parameters (this number $(k+1)(k+2)/2$ for scalar equations).
\vfill \eject
\noindent 4. \souligne {\bf {Example}}

In this section, we discuss a system of two coupled
equations which admits algebraic solutions. This example
arises in the study of the stability about sphalerons [7]
(i.e. unstable classical solutions)
in the Abelian gauge-Higgs model in 1+1 dimension [8].
The relevant  Schr\"odinger equation reads
$$\pmatrix { {d^2 \over dz^2} + \lambda - \theta^2k^2{\rm {sn}}^2
 &-2\theta k\ {\rm {cn \cdot dn}} \cr
-2 \theta k\ {\rm {cn \cdot dn}}
& {d^2\over dz^2} +
\lambda +1+k^2 -(\theta^2+2)k^2{\rm {sn}}^2\cr} \pmatrix {f(z)\cr
g(z)\cr} = 0\eqno(33)$$
\noindent
and is considered on the Hilbert space of
periodic functions over [0,4K(k)]  (K(k) is
the complete elliptic integral of the second type).
The three elliptic functions [9]
 ${\rm {sn}}(z,k), {\rm {cn}}(z,k), {\rm {dn}}(z,k)$ are
periodic with periods 4K(k), 4K(k), 2K(k) respectively.
The spectral parameter  $\lambda$ represents the mode eigenvalue:
negative $\lambda$'s correspond to the
unstable modes of the sphalerons.

The parameter  $\theta$ depends on the coupling constants of the model;
it represents the mass ratio $2M_H/M_W$
where $M_W$ (resp. $M_H$) is the mass of the gauge (resp. Higgs) boson.
The system above admits algebraic solutions if these masses
are such that
$$  \theta^2 = N(N+1) \ \ \ {\rm or} \ \ \ \ M_H^2 = {N(N+1) \over 4} M_W^2
  \ \ \ {\rm with } \ \ N\ {\rm integer} \eqno(34)$$
Following a similar scenario as for the Lam\'e equation [10]
$$  ( {d^2\over dz^2} + \lambda  - N(N+1) k^2{\rm {sn}}^2) f(z) = 0
\eqno(35)$$
the  algebraic solutions of Eq.(33) occurs in four sectors of the
Hilbert space [11].
In order to construct them, we need to set the
system in the forms discussed previously.
For this purpose, we perform the change of variable
$$  x = {\rm sn}(z,k)^2 \eqno(36) $$
accompanied by a change to new functions $P(x), Q(x)$ defined through
 $$    f(z) =  F(z) P(x) \ \ \ \ , \ \ \ g(z) = G(z) Q(x) \eqno(37) $$
where $F(z), G(z)$  are some products of  ${\rm sn, cn, dn}$.
Then we have to determine $F(z)$ and $G(z)$ so that the new
equations for $P(x), Q(x)$ admit polynomial solutions in the variable $x$.
The factors $F(z), G(z)$ allowing for an algebraisation are different
according to the parity of $N$; the eight different possibilities
are reported  in  the first two columns of Table 1.

Even if the changes of variable (36)  and of function (37) are performed,
the equation for $P, Q$ is still not  in a form that preserves
${\cal P}_{m,n}$: an additional change of basis of the form
$$ \pmatrix { \tilde P(x) \cr \tilde Q(x) \cr} =
         V^{-1}  \pmatrix { P(x) \cr  Q(x) \cr}  \eqno (38)  $$
need to be done. In the eight cases, the change of basis $V$ can be
constructed from the following matrices:
$$ \alpha = \pmatrix {1 &0 \cr
                      0  &{\theta \over N} \cr} \ \ \ , \ \ \
   \beta(k)  =  \pmatrix {1 &0 \cr
                       0  &k \cr} \ \ \ , \ \ \
   \gamma(x) =   \pmatrix {1 &x \cr
                           0  &1 \cr} \ \ \ , \ \ \
   \delta =   \pmatrix {1 &1+N \cr
                        -1&N \cr}   \eqno(39)$$
They  are presented in the third column of Table 1 and
the dimensions $m,n$ of the  invariant space
${\cal P}_{m,n}$ are reported in the last two columns.
these results demonstrate that Eq.(33) admits four types
of algebraisation, yielding a total of $4N+2$ algebraic solutions,
 if $N$ is an integer (this number is $2N+1$ for the Lam\'e equation).
The first (resp. last) four lines in  the table correspond
to the solutions available for $N$ odd (resp. even).


 The analogy between Eqs. (33) and (35)
is also present when doubly periodic solutions [10]
are considered,  i.e. solutions of period 8K(k).
In order to discuss this issue we  consider the following
changes of functions:

$$ \eqalign {
   &f(z) = \sqrt{ {\rm dn}(z) \pm {\rm cn}(z) }
                       (\pm Y(x){\rm cn}(z) + Z(x){\rm dn}(z)  )\cr
   &g(z) = \sqrt{ {\rm dn}(z) \pm {\rm cn}(z) }
                       (\pm V(x){\rm cn}(z) + W(x){\rm dn}(z)  )\cr
          } \eqno(40)$$
and
$$ \eqalign {
   &f(z) = \sqrt{ {\rm dn}(z) \pm {\rm cn}(z) }
                       (\pm Y(x){\rm cn}(z) {\rm dn}(z) + Z(x) )\cr
   &g(z) = \sqrt{ {\rm dn}(z) \pm {\rm cn}(z) }
                       (\pm V(x){\rm cn}(z) {\rm dn}(z) + W(x) )\cr
          } \eqno(41)$$
Due to the square root factor, these functions have 8K(k) as  period,
the functions with $+$ and $-$ correspond to each other
by a translation $z \rightarrow z + 2K(k)$.

Inserting the ansatzes (40) into Eq.(33) and
identifying to zero the coefficients of the  factors
cn, dn,  yields a system of four equations in $Y,Z,V,W$.
To put this system into a cannonical form; we must
perfom an additional change of basis. In this case, it reads
$$ \pmatrix {\tilde Y \cr \tilde W \cr \tilde Z \cr  \tilde V \cr }
=  \pmatrix { 1         &1          &0          &0 \cr
             {k \theta \over N+1} &{-k \theta \over N} &0 &0 \cr
              0         &0          &1          &1 \cr
              0         &0    &{k \theta \over N+1} &{-k \theta \over N}\cr}
 \pmatrix { Y \cr  W \cr  Z \cr  V \cr } ;
\eqno(42)$$
then the differential operator associated with the ansatz (40)
preserves the space (using obvious notations)
$$ {\cal P}_{ \tilde N-1, \tilde N, \tilde N-1, \tilde N }
\ \ \ \ {\rm if} \ \ \
         N = {1 \over 2} + 2 \tilde N,   \ \  \ \tilde N \ \ {\rm integer}
\eqno(43)$$

Similarly, after the change of basis defined by
$$ \pmatrix {\tilde Y \cr \tilde W \cr \tilde Z \cr  \tilde V \cr }
=  \pmatrix { 1         &0          &0          &0 \cr
            {k \theta \over N} x  &1  &0        &0 \cr
              0         &0          &1          &{k \theta \over N+1} x \cr
              0         &0          &0          &1 \cr}
 \pmatrix { Y \cr  W \cr  Z \cr  V \cr }
\eqno(44)$$
we obtain the operator associated with the ansatz (41) in a form that
preserves the space
$$ {\cal P}_{ \tilde N, \tilde N, \tilde N, \tilde N }
 \ \ \ \ {\rm if} \ \ \
         N = {3 \over 2} + 2 \tilde N,  \ \ \ \tilde N \ \ {\rm  integer}
\eqno(45)$$
This demonstrates that $4N+2$ algebraic solutions are available also
when $N$ is a semi-integer once the  degeneracy $\pm$
in Eqs.(40,41) is taken into account
(both signs yield solutions of equal eigenvalues).


\noindent 5. \souligne {Conclusions}

Quasi-exactly-solvable equations constitute an attactive
bridge between group theory and spectral equations.
While ordinary QES equations are related to usual
Lie algebra, systems of coupled QES equations are related to
more sophisticated structures.
For systems of two coupled equations, the relevant operators are
those that preserve the spaces ${\cal P}_{n,m}$.
These operators can be perceived as projectivised
representations of some graded algebra.
For our equations, the relevant  graded algebra have
finite numbers of generators and our results demonstrate that they
admit representations of arbitrary finite dimension.
The anti-commutators between the fermionic generators
close into $n-m$-powers of the bosonic generators.

Applications of quasi exactly solvable systems
can be found in the framework of  quantum
mechanics, namely in the study of coupled channels.
Here, we illustrate the relevance  of  such systems  within another
domain, namely in the study of unstable modes about sphalerons
in a simple field theory. This example suggests that the stability
analysis about  other classical solutions (solitons, kinks....) might
also be related to QES systems (see, however, Ref.[12]).

\vfill \eject

\noindent {\bf  REFERENCES}
\item {[1]} A.V. Turbiner, Comm. Math. Phys. 118 (1988) 467.
\item {[2]} A.G. Ushveridze, Fiz. Elem. Chast. Atom. Yad,
            20 (1989) 1185, [Sov. Journ. Part. Phys. 20 (1989) 504].
\item {[3]} M.A. Shifman, Int. Journ. Mod. Phys. A4 (1989) 2897.
\item {[4]} M.A. Shifman and A.V. Turbiner, Comm. Math. Phys. 120 (1989)347.
\item {[5]} A.V. Turbiner, Lie-algebraic approach to the theory of
polynomial solutions. I. Ordinary differential equations
and finite-difference equations in one
variable. CPT-91/P.2628.
\item {[6]} A.V. Turbiner, Lie-algebraic approach to the theory of
polynomial solutions. II. Differential equations in one real and one Grassmann
variables and $2\times 2$ matrix differential equations. ETH-TH/92-21.
\item {[7]} F. R. Klinkhamer and N. S. Manton, Phys. Rev. D30 (1984).
\item {[8]} Y. Brihaye, P. Kosinski, S. Giller and
            J. Kunz, Phys. Lett. B293 (1992) 383.
\item {[9]} I.S. Gradshtein and I.M. Ryzhik, Table of integrals,
series and products (Academic Press, New York, 1965).
\item {[10]} F.M. Arscott, Periodic differential equations (Pergamon,
Oxford, 1964).
\item {[11]} A.V. Turbiner, J. Phys. A 22 (1989) L1.
\item {[12]} D.P. Jatkar, C.N. Kumar and A. Khare, Phys. Lett. A142
(1989) 200.
\vfill \eject

\centerline {\souligne {\bf {TABLE 1}}}
$$\vbox{\tabskip=0pt\offinterlineskip
\def\tablerule{\noalign{\hrule}}
\halign to 16 true cm{\strut #& \vrule#\tabskip=1em plus 2em&
\quad\hfil#\hfil&\vrule#&
\quad\hfil#\hfil&\vrule#&
\quad\hfil#\hfil&\vrule#&
\quad\hfil#\hfil&\vrule#&
\quad\hfil#\hfil&\vrule#\tabskip=0pt\cr\tablerule
&
&\omit\hidewidth $F(z)$   \hidewidth&
&\omit\hidewidth $G(z)$   \hidewidth&
&\omit\hidewidth $V$      \hidewidth&
&\omit\hidewidth $m$      \hidewidth&
&\omit\hidewidth $n$      \hidewidth&\cr\tablerule
&&\ & &\ & &\ & &\ & &\ &\cr
&&   1& &${\rm cn\ dn}$& &$\alpha\ \beta(1/k)\ \gamma$& &${N-1 \over 2}$&
                                                     &${N-1 \over 2}$&\cr
&&\ & &\ & &\ & &\ & &\ &\cr
&&${\rm cn\ dn}$& &1& &$\alpha\ \beta(k)\ \gamma^t$& &${N-1 \over 2}$&
                                                      &${N-1 \over 2}$&\cr
&&\ & &\ & &\ & &\ & &\ &\cr
&&${\rm sn\ cn}$& &${\rm sn\ dn}$& &$\alpha\ \beta(1/k)\ \delta$& &${N-3 \over 2}$
                                                    &${N-1 \over 2}$&\cr
&&\ & &\ & &\ & &\ & &\ &\cr
&&${\rm sn\ dn}$& &${\rm cn\ cn}$& &$\alpha\ \beta(k)\ \delta$& &${N-3 \over
2}$
                                                    &${N-1 \over 2}$&\cr
&&\ & &\ & &\ & &\ & &\ &\cr
&&${\rm sn}$& &${\rm sn\ cn\ dn}$& &$\alpha\ \beta(1/k)\ \gamma$& &${N-2 \over
2}$
                                                     &${N-2 \over 2}$&\cr
&&\ & &\ & &\ & &\ & &\ &\cr
&&${\rm sn\ cn\ dn}$& &${\rm sn}$& &$\alpha\ \beta(k)\
                                 \gamma^t$& &${N-2\over 2}$&
                                                  &${N-2 \over 2}$&\cr
&&\ & &\ & &\ & &\ & &\ &\cr
&&${\rm cn}$& &${\rm  dn}$& &$\alpha\ \beta(1/k)\ \delta$& &${N-2 \over 2}$&
                                                    &${N \over 2}$&\cr
&&\ & &\ & &\ & &\ & &\ &\cr
&&${\rm dn}$& &${\rm cn}$& &$\alpha\ \beta(k)\ \delta$& &${N-2 \over 2}$&
                                                    &${N \over 2}$&\cr
&&\ & &\ & &\ & &\ & &\ &\cr
\tablerule\cr}}$$

\noindent The factors $F(z),G(z)$ allowing for an algebraisation
of the system (33) are given in the first two colums. The third
column contains the change of basis (38) and the last one
the degree of the invariant polynomial space ${\cal P}_{m,n}$

\vfill \end


\vfill \end

Here  HeApart from their interrest in quantum mechanics,

\par In the first part we have characterized QES equations that admit more
algebraic solutions than expected by the general theory.
We have shown that the occurence of these  multiple algebraizations is
related to the fact that some coefficients
appearing in the equations  are integer or not.
Concrete examples of such equations are discussed in Refs.[8,13].
We also establish a connection between some non QES equations
and systems of coupled QES equations.
In particular, we have demonstrated that
the (non polynomial) Lam\'e solutions can be perceived in this way.
\par

Several kinds of  algebraic structures were studied recently  in
different contexts of theoretical physics:
graded algebra, Krishever-Novikov algebra,  W-algebra.
 The relevance of QES systems
in conformal field theories  was pointed out in Ref.[14].

\vfill\eject
\item {[8]} Y. Brihaye and S. Braibant, Quasi-exactly-solvable
            systems and sphaleron stability, Journ. Math. Phys. (to appear).
\item {[9]} F.M. Arscott, Periodic differential equations (Pergamon,
Oxford, 1964).
\item {[12]} Y. Brihaye (in preparation).
\item {[13]} J.-Q. Liang, H.J.W. Mller-Kirsten and D.H. Tchrakian,
 Phys. Lett. B282 (1992) 105.
\item {[14]} A. Morozov, A. M. Perelomov, A.A. Rosly,
M.A. Shifman and A.V. Turbiner,\hfill\break
Int. Journ.Mod.Phys. A5 (1990) 803.

\vfill\eject

{\souligne {\bf {Appendix A}}}: Some explicit solutions

This Appendix contains a few explicit solutions of the
equations studied above.

\noindent 1. Solutions of Eq.(5) for $n=0$ i.e. $\delta=0$
$$  \lambda = 0 \ \ \ \ , \ \ \ \ G = 1  \eqno (A1)$$

\noindent  2. Solutions of Eq.(5) for $n=1$
i.e. $\delta=6 + 2(\gamma - \alpha - \beta)$
$$ \eqalign{
            \lambda &= (\alpha - \gamma -2)k^2
             +(\beta - \gamma -2)   \pm \sqrt{\Theta} \cr
            \Theta &= k^4(\alpha - \gamma - 2)^2
                     +2 k^2(\alpha \beta + \alpha \gamma + \beta \gamma
                             - (\gamma + 2)^2 + 2)
                     + (\beta - \gamma -2)^2 \cr
             G(t) &=  2 k^2 (\alpha + \beta - \gamma - 3) t
                 - k^2( \alpha - \gamma - 2) - (\beta - \gamma - 2)
                   \pm \sqrt{\Theta} \cr
           }  \eqno (A2)$$
In each case, seven other solutions follows from the substitutions (12-16).

\noindent 3.  Solutions of Eq.(25) for $n=0$ i.e. $\Delta=0$
$$  \eqalign{
      & \Lambda = (\mu + 1)(1+k^2) \pm
                 \sqrt{(\mu+1)^2 (1+k^2)^2 - 4 k^2((\mu+1)^2 - \nu^2)} \cr
      & Y(t) = \nu k^2 \ \ \ \ , \ \ \ \
        Z(t) = \Lambda - \mu - 1 \cr } \eqno (A3)$$

\noindent 4.  Solutions of Eq.(25) for $n=0$ i.e. $\Delta=\mu+2$
$$  \Lambda = 0 \ \ \ \ ,
            \ \ \ \ Y(t) = 1 \ \ \ \ , \ \ \ \ Z(t) = 0   \eqno (A4)$$
\vfill \eject
{\souligne {\bf {Appendix B}}}: The algebra preserving ${\cal P}_{n,n-2}$

In this appendix we give the list of (anti-)commutation
relations fullfilled by the 10 operators (33).

$$ [ T^+ , T^-] = -2 T^0 \ \ \ , \ \ \
   [ T^{\pm} , T^0] = \mp  T^{\pm}  \ \ \ , \ \ \
   [ T^a , J] = 0 \eqno (A1)$$

$$ \eqalign {
  &[Q_1 , T^-] = 0 \ \ \ , \ \ \
   [Q_1 , T^0] = Q_1 \ \ \ , \ \ \
   [Q_1 , T^+] = 2 Q_2 \ \ \ , \ \ \ \cr
  &[Q_2 , T^-] = -Q_1 \ \ \ , \ \ \
   [Q_2 , T^0] = 0 \ \ \ , \ \ \
   [Q_2 , T^+] = Q_3 \ \ \ , \ \ \ \cr
  &[Q_3 , T^-] = -2 Q_2 \ \ \ , \ \ \
   [Q_3 , T^0] = -Q_3 \ \ \ , \ \ \
   [Q_3 , T^+] = 0 \ \ \ , \ \ \
  \cr }  \eqno (A2)$$

$$ \eqalign {
  &[\bar Q_1 , T^-] = 0 \ \ \ , \ \ \
   [\bar Q_1 , T^0] = \bar Q_3 \ \ \ , \ \ \
   [\bar Q_1 , T^+] = 2 \bar Q_2 \ \ \ , \ \ \ \cr
  &[\bar Q_2 , T^-] = -\bar Q_3 \ \ \ , \ \ \
   [\bar Q_2 , T^0] = 0 \ \ \ , \ \ \
   [\bar Q_2 , T^+] = \bar Q_1 \ \ \ , \ \ \ \cr
  &[\bar Q_3 , T^-] = -2 \bar Q_2 \ \ \ , \ \ \
   [\bar Q_3 , T^0] = -\bar Q_1 \ \ \ , \ \ \
   [\bar Q_3 , T^+] = 0 \ \ \ , \ \ \
  \cr }  \eqno (A3)$$

$$ \eqalign {
  &\{ Q_1 , \bar Q_3 \} = (T^-)^2    \cr
  &\{ Q_1 , \bar Q_2 \} = (T^0 + J_c) T^- \ \ \ ,\ \ \
   \{ Q_2 , \bar Q_3 \} = (T^0 + J  ) T^-       \cr
  &\{ Q_1 , \bar Q_1 \} =  (T^0-J)(T^0+J_c)	\cr
  &\{ Q_2 , \bar Q_2 \} = {1\over 2} (T^+T^- + (T^0+J)(T^0+J_c)) \cr
  &\{ Q_3 , \bar Q_3 \} = (T^0+J)(T^0-J_c)	\cr
  &\{ Q_2 , \bar Q_1 \} = T^+ (T^0 + J_c) \ \ \ , \ \ \
   \{ Q_3 , \bar Q_2 \} = T^+ (T^0 + J)   	\cr
  &\{ Q_3 , \bar Q_1 \} = (T^+)^2
  \cr } \eqno (A.4)$$
where for convenience we defined
$$  J_c \equiv \1op - J  \eqno (A5)$$

\vfill \eject
\tablerule\cr}}$$
\vfill \end

\item {[7]} A. I. Bochkarev and M. E. Shaposhnikov, Mod. Phys. Lett.
            A2 (1987) 991.
\item {[8]} L. Carson, Phys. Rev. D42 (1990) 2853.
\item {[10]} N. S. Manton and T. N. Samols, Phys. Lett. B207 (1988) 179.
\item {[11]} J. Kunz and Y. Brihaye, Phys. Lett. B216 (1989) 353.
\item {[12]} L. G. Yaffe, Phys. Rev. D40 (1989) 3463.
\item {[14]} J.-Q. Liang, H.J.W. Mller-Kirsten and D.H. Tchrakian,


\magnification 1200
\baselineskip=16pt
\input frtex
\nopagenumbers
\headline={\hfill \tenrm\folio\ }
\parskip 10pt plus 1pt minus 1pt
\hsize 16 true cm \vsize 24 true cm
\def\relatif{\ \hbox{{\rm Z}\kern-.4em\hbox{\rm Z}}}
\def\reel{\ \hbox{{\rm R}\kern-1em\hbox{{\rm I}}}}
\def\souligne#1{$\underline{\smash{\hbox{#1}}}$}
\def\1op{\hbox{{\rm 1}\kern -0.23em\hbox{{\rm I}}}}
\def\laeq{\ \raise .3ex\hbox{{\rm {$<$}}\kern-.8em\lower 1.3ex\hbox{{\rm
{$\sim$}}}}\ }

\vfill\end

\noindent {\bf  REFERENCES}
\item {[1]} A.V. Turbiner, Comm. Math. Phys. 118 (1988) 467.
\item {[2]} A.G. Ushveridze, Fiz. Elem. Chast. Atom. Yad,
            20 (1989) 1185, [Sov. Journ. Part. Phys. 20 (1989) 504].
\item {[3]} A.V. Turbiner, J. Phys. A 22 (1989) L1.
\item {[4]} M.A. Shifman, Int. Journ. Mod. Phys. A4 (1989) 2897.
\item {[5]} Y. Brihaye, P. Kosinski, S. Giller and
            J. Kunz, Phys. Lett. B293 (1992) 383.
\item {[6]} F. R. Klinkhamer and N. S. Manton, Phys. Rev. D30 (1984).
\item {[7]} A. I. Bochkarev and M. E. Shaposhnikov, Mod. Phys. Lett.
            A2 (1987) 991.
\item {[8]} L. Carson, Phys. Rev. D42 (1990) 2853.
\item {[9]} I.S. Gradshtein and I.M. Ryzhik, Table of integrals,
series and products (Academic Press, New York, 1965).
\item {[10]} N. S. Manton and T. N. Samols, Phys. Lett. B207 (1988) 179.
\item {[11]} J. Kunz and Y. Brihaye, Phys. Lett. B216 (1989) 353.
\item {[12]} L. G. Yaffe, Phys. Rev. D40 (1989) 3463.
\item {[13]} F.M. Arscott, Periodic differential equations (Pergamon,
Oxford, 1964).
\item {[14]} J.-Q. Liang, H.J.W. Mller-Kirsten and D.H. Tchrakian,
Phys. Lett. B282 (1992) 105.

\vfill\eject
\noindent {\bf {APPENDIX A}}
\vskip 0.5 true cm
\noindent In this Appendix, we give the solutions of Eq.(3)
for $N=1$ and $N=2$. The  Lam\'e polynomials (i.e. the
non trivial values of $W$) are normalized as in Ref. [9].

\noindent  {\bf N = 1}
$$\eqalign {
a)\ &W=0,\ \ \ \ \ \ \
h=-(1+k^2), \ \ \
\ \ \ F ={\rm {cn \cdot dn}}\cr
b)\ &W=0,\ \ \ \ \ \ \
h=-2k,\ \ \ \ \ \ \ \ \
\ \ \ F = k^2{\rm {sn}}^2- k\cr
c)\ &W=0,\ \ \ \ \ \ \
h=2k,\ \ \ \ \ \ \ \ \ \
\ \ \ F = k^2{\rm {sn}}^2+ k\cr
d)\ &W={\rm {sn}}, \ \ \
h=1+k^2,\ \ \
\ \ \ F =-{\rm {cn}} \cdot {\rm {dn}}/h \cr
e)\ &W={\rm {cn}}, \ \ \
h=1,\ \ \ \ \ \ \ \
\ \ \ F = {\rm {sn}} \cdot {\rm {dn}}\cr
f)\ &W={\rm {dn}}, \ \ \
h=k^2,\ \ \ \ \ \
\ \ \ F = {\rm {sn}} \cdot {\rm {cn}}\cr} \eqno (A1)$$

\noindent  {\bf N = 2}
$$\eqalign {
a)\ &W=0,\ \ \
h=(1-2\sqrt {1+3k^2}), \ \ \ \ \
F = {\rm {cn}} \cdot (6k^2{\rm {sn}}^2+h-3)\cr
b)\ &W=0,\ \ \
h=k^2(1-{2\over {k}}\sqrt {3+k^2}),\ \ \
F = {\rm {dn}} \cdot (6k^2{\rm {sn}}^2+h-3k^2)\cr
c)\ &W=0,\ \ \
h=k^2(1+{2\over {k}}\sqrt {3+k^2}),\ \ \
F = {\rm {dn}} \cdot (6k^2{\rm {sn}}^2+h-3k^2)\cr
d)\ &W=0,\ \ \
h=1+2\sqrt {1+3 k^2},\ \ \
F = {\rm {cn}} \cdot (6k^2{\rm {sn}}^2+h-3)\cr
e)\ &W=0,\ \ \
h=3(1+k^2),\ \ \ \ \ \ \ \
F = k^3{\rm {sn}}^3 \cr
f)\ &W={\rm {sn}} \cdot {\rm {cn}},\ \ \
h=4+k^2,\ \ \
\ \ \ F = {\rm {dn}}\cdot (2{\rm {sn}}^2-1)/h\cr
g)\ &W={\rm {sn}} \cdot {\rm {dn}},\ \ \
h=1+4k^2,\ \ \
\ \ \ F = {\rm {cn}} \cdot (2k^2{\rm {sn}}^2-1)/h\cr
h)\ &W={\rm {cn}} \cdot {\rm {dn}},\ \ \
h=1+k^2,\ \ \
\ \ \ F = {\rm {sn}} \cdot (1+k^2-2k^2{\rm {sn}}^2)/h\cr
i)\ &W={\rm {sn}}^2 -{ h_{\mp} \over 6k^2},\ \ \
h_{\pm} = 2(1+k^2) \pm 2\sqrt {1-k^2+k^4},\ \ \
\ \ \  F = -2 {\rm {sn \cdot cn \cdot dn}} /
h_{\pm}\cr} \eqno(A2)$$
   \vfill\eject
\noindent {\bf {APPENDIX B}}
\vskip 0.5 true cm
\noindent
In this Appendix, we give the solutions of Eq.(3)
for $N=1/2$ and $N=3/2$. For each solution,
there is another one  with the same eigenvalue
and period; it is obtained by changing the sign of the function cn
everywhere (i.e. in  $W$ and $F$).

\noindent {\bf N = 1/2}
$$\eqalign {
a)\ &W=0\ \ \ ,\ \ \
 h=-{3\over 4} (1+k^2)\ \ \ ,\ \ \
 F=({\rm {dn}}+{\rm {cn}})^{3\over 2}\ \ \cr
b)\ &W=({\rm {dn-cn}})^{1\over 2}\ \ \ ,\ \ \
 h={1\over 4} (1+k^2)\ \ ,\ \
 F= {2\sqrt{1-k^2} \over {1-k^4}}
 ({\rm {dn}}+{\rm {cn}})^{1\over 2}(k^2{\rm {cn}}-{\rm {dn}})\ \
\cr} \eqno (B1)  $$

\noindent {\bf N = 3/2}
$$\eqalign{
a) \ &W=0\ \  , \ \
h={1+k^2\over 4} \pm \sqrt {1+7k^2+k^4}\ \ , \cr
&\ \ \ \ F=({\rm {dn}}+{\rm {cn}})^{3\over 2} \lbrace (7k^2+3+4h){\rm {cn}}
-(3k^2+7+4h){\rm {dn}}\rbrace \cr
b) \ &W = ({\rm {dn}}-{\rm {cn}})^{1\over 2}
      ({\rm {dn}}-(k^2-1+\Delta){\rm {cn}})\ \ ,
\ \ h = {5\over 4} (1+k^2)-\Delta\ \ , \ \
\Delta \equiv \pm \sqrt {1-k^2+k^4} \cr
&\ \ \ \ F =C({\rm {dn}}+{\rm {cn}})^{1\over 2} \lbrace 3k^2{\rm
{sn}}^2+3(k^2+1-\Delta){\rm {cn.dn}}-(2k^2+2-\Delta)\rbrace \cr
&\ \ \ \  C={2\over 3} {(9k^4+k^2+4)+\Delta(9k^2+5)\over {3k^4 +22k^2+3}}
} \eqno (B2)$$
\vfill\end
\magnification 1300
\input frtex
\voffset 0.7 true cm
\baselineskip=14pt
\hsize 16.5 true cm
\vsize 24 true cm

$$\eqalign {\lbrace \bar Q_{i\alpha}, Q_{j\beta}\rbrace
&= 2 T_i(T-\sigma)_{\alpha\beta} T_j - {2\over 3} T^2\delta_{ij}
(T-\sigma)_{\alpha\beta}\cr
&+ {i\over 3} \epsilon_{ijk} T_k T^2 \delta_{\alpha\beta}-{1\over 3}
T^2(T_i\sigma_j+T_j\sigma_i)_{\alpha\beta}\cr
&+ 2(J-1) (T_iT_j-{2\over 3} \delta_{ij}T^2)\delta_{\alpha\beta}\cr
&+ i(J-{3\over 2})\epsilon_{ijk} \lbrace T_k, T_{\ell}\rbrace
(\sigma_{\ell})_{\alpha\beta}\cr
&+{i\over 2} (\lbrace T_j, T_k\rbrace \epsilon_{ik\ell}-\lbrace T_i,
T_k\rbrace \epsilon_{jkl}(\sigma_{\ell})_{\alpha\beta}\cr
&-{i\over 3} (J-1)\epsilon_{ijk} T^2(\sigma_k)_{\alpha\beta}\cr
&+{i\over 3} (J-2) (5J-3)\epsilon_{ijk} T_k \delta_{\alpha\beta}\cr
&-{4\over 3} J(J-2) \delta_{ij}(T_k\sigma_k)_{\alpha\beta}\cr
&+{1\over 3} J(J-2) \lbrace
T_k(\sigma_j)_{\alpha\beta}+T_j(\sigma_i)_{\alpha\beta}\rbrace\cr
&+{i\over 3} J(J-1)(J-2) \epsilon_{ijk} (\sigma_k)_{\alpha\beta}-{2\over
3} J(J-1)J-2)\delta_{ij}\delta_{\alpha\beta}\cr}$$
\vskip 1 true cm
$$T^3 = T^0\ \ , \ \ T^1 = {1\over 2} (T^--T^+)\ \ ,\ \ T^2  = {i\over 2
} (T^-+T^+)$$
\vskip 1 true cm
$$\eqalign {
&Q_{3{1\over 2}} = \sqrt {2} Q_3\ \ \ \bar Q_{3{1\over 2}} = \sqrt {2}
\bar Q_3\cr
&Q_{3{1\over 2}} = \sqrt {2} Q_2\ \ \ \bar Q_{3-{1\over 2}} = -\sqrt {2}
\bar Q_2\cr
&Q_{1{1\over 2}} = {1\over {\sqrt {2}}} (Q_2-Q_1) \ \ \
\bar Q_{1-{1\over 2}} = -{1\over {\sqrt {2}}}
(\bar Q_2-\bar Q_4)\cr
&Q_{1-{1\over 2}} = {1\over {\sqrt {2}}} (Q_1-Q_3) \ \ \
\bar Q_{1-{1\over 2}} = {1\over {\sqrt {2}}}
(\bar Q_1-\bar Q_3)\cr
&Q_{2-{1\over 2}} = {i\over {\sqrt {2}}} (Q_2+Q_4) \ \ \
\bar Q_{2{1\over 2}} = {+i\over {\sqrt {2}}}
(\bar Q_2-\bar Q_4)\cr
&Q_{2-{1\over 2}} = {i\over {\sqrt {2}}} (Q_1+Q_3) \ \ \
\bar Q_{2-{1\over 2}} = {-i\over {\sqrt {2}}}
(\bar Q_1-\bar Q_3)\cr
&Q_{i\alpha}(\sigma_i)_{\alpha\beta} = 0\ \ \ ,\ \ (\sigma_i)_{\alpha
\beta} \bar Q_{i\beta} = 0\cr
&[T_i,Q_{j\alpha}] = i\epsilon_{ijk}Q_{k\alpha} + {1\over 2}
Q_{j\beta} (\sigma_i)_{\beta\alpha}\cr
&[T_i, \bar Q_{j\alpha}] = i\epsilon_{ijk} \bar Q_{k\alpha} - {1\over 2
} (\sigma_i)_{\alpha\beta} \bar Q_{j\beta}\cr}$$

\vfill \end
